\documentclass[preprint]{revtex4}
\usepackage{graphicx}
 
\begin{document}

\title[Frequency doubling of temporally incoherent light from a superluminescent diode]{Frequency doubling of temporally incoherent light from a superluminescent diode in a periodically poled lithium niobate waveguide crystal}

\author{Henning Kurzke, Jan Kiethe, Axel Heuer and Andreas Jechow}

\address{University of Potsdam, Institute of Physics and Astronomy, Photonics, Karl Liebknecht Str. 24-25, 14476 Potsdam}

\begin{abstract} The amplified spontaneous emission from a superluminescent diode was frequency doubled in a periodically poled lithium niobate waveguide crystal. The temporally incoherent radiation of such a superluminescent diode is characterized by a relatively broad spectral bandwidth and thermal-like photon statistics, as the measured degree of second order coherence, g$^{(2)}$(0)=1.9$\pm$0.1, indicates. Despite the non-optimized scenario in the spectral domain, we achieve six orders of magnitude higher conversion efficiency than previously reported with truly incoherent light. This is possible by using single spatial mode radiation and quasi phase matched material with a waveguide architecture. This work is a principle step towards efficient frequency conversion of temporally incoherent radiation in one spatial mode to access wavelengths where no radiation from superluminescent diodes is available, especially with tailored quasi phase matched crystals. The frequency doubled light might find use in applications and quantum optics experiments.\end{abstract}


\maketitle
\section*{Introduction}
Incoherent or thermal light is ubiquitous in nature. Modern applications of temporally incoherent (but usually spatial single mode) light include speckle reduction, optical coherence tomography \cite{huang1991OCT}, ghost imaging \cite{Gatti2004}, sub-wavelength lithography \cite{Cao2010}, metrology \cite{Zhu2012} and repumping of laser cooled ions from a dark state \cite{Lindvall2013}. Some of these applications \cite{Gatti2004, Cao2010, Zhu2012} exploit the quantum nature of light via the photon statistics, which can be determined by the degree of second order coherence (DSOC) g$^{(2)}$(0). Light with thermal-like photon statistics will yield a DSOC g$^{(2)}$(0) of 2 which is called photon bunching \cite{HBT1956}, while coherent light will give a value of 1. Due to the short coherence times of thermal-like radiation a measurement of the DSOC could not be demonstrated until recently. Boitier et al. \cite{Boitier2009} measured the DSOC of the amplified spontaneous emission (ASE) from a fiber source to be 1.8$\pm$0.1 with fs resolution using two-photon absorption in a photo-multiplier tube (PMT). Since then, several groups have studied thermal-like light emitted from ASE sources including superluminescent diodes (SLDs) \cite{Blazek2011}. We have recently utilized photon bunching in ASE from such an SLD to demonstrate enhanced two-photon excited fluorescence, which has potential application in microscopy \cite{jechow2013np}.

Second harmonic generation (SHG) is a standard way to reach wavelength regions were no other light sources are available. Typically, lasers are used because of the stringent phase matching conditions of the nonlinear materials. However, shortly after the first demonstration of SHG with laser in 1961 \cite{franken1961} several attempts were made to use incoherent light sources for nonlinear frequency conversion \cite{mcmahon1965, mcmahon1966}. Due to the low efficiency of 10$^{-14}$ of the nonlinear process the approach to frequency double incoherent light was omitted in favor to the much more efficient frequency doubling of coherent radiation. Recently, the study of SHG with truly incoherent light sources was revoked in seminal work with sunlight \cite{tamosauskas2011} and a tungsten lamp \cite{stabinis2011spectrum}. Unfortunately, the conversion efficiency was still low reaching values on the order of 10$^{-11}$. However, since the availability of highly efficient nonlinear materials exploiting quasi phase matching (QPM) as bulk and waveguide crystals, it is feasible to convert radiation with low intensities e.g. emission from Silicon-vacancy (SiV) centers in diamond \cite{Zaske2011} or squeezed light \cite{vollmer2014} with the help of coherent radiation in optical parametric amplifiers (OPAs). Furthermore, these engineered QPM waveguides serve as efficient sources for entangled photons by spontaneous parametric down conversion \cite{Jechow2008, Eckstein2011}.

Here we utilize ASE emitted by an SLD for SHG in a periodically poled lithium niobate (PPLN) waveguide crystal and achieve conversion efficiencies on the order of 10$^{-5}$, which is six orders of magnitude higher than previously reported. We determine the DSOC function g$^{(2)}$(0) of the ASE to be 1.9$\pm$0.1, validating the thermal-like photon statistics of the light source and therefore its temporal incoherence. Despite the non optimized scenario, our results show that efficient nonlinear frequency conversion of thermal-like light without the aid of coherent radiation can be achieved. By tailoring the poling period of the nonlinear crystal with e.g. a chirped grating it should be feasible to further improve the overall efficiency of the process resulting in broadband upconverted light. Our scheme might be of interest for upconversion imaging \cite{Dam2010} as well as photovoltaics where upconversion crystals \cite{zou2012} or triplet-triplet annihilation \cite{Baluschev2006} systems are used to shift the wavelengths of sunlight to match the band gap of solar cells. Furthermore, SHG of thermal-like light with g$^{(2)}$(0)=2 might result in super-bunched light with g$^{(2)}$(0)=6 \cite{loudon}, which is of high interest for aforementioned applications and quantum optics research.

\section*{Experimental setup}
The experimental setup is depicted in Fig. 1. The emission from an SLD is coupled into a polarization maintaining single mode fiber by using two aspherical lenses( L1, f$\,$=$\,$8$\,$mm, L2 f$\,$=$\,$7.5$\,$mm). The light exiting the fiber is collimated with an aspherical lens L3 with a focal length of 8$\,$mm and coupled into the PPLN waveguide crystal by an aspherical lens (L4 f$\,$=$\,$8$\,$mm). The frequency doubled light is recollimated by an aspherical lens (L5 f$\,$=$\,$11$\,$mm) and separated from the infrared pump light by a short pass filter and a color glass.

The SLD was manufactured by \textsc{m2k laser} and has an emitter width of $w$~=~400~$\mu$m and a chip length of $l$~=~1500~$\mu$m. To suppress laser activity the front facet was AR coated providing a residual reflectivity of $R_{front} \approx 4\cdot 10^{-4}$, while the back facet had a reflectivity of $R_{back} >$~90~\%. Typically, these devices were used in an external cavity setup to achieve high brightness emission \cite{Jechow2009, Heuer2012}.

The nonlinear crystal was manufactured by \textsc{HCPhotonics} and has a length of 10$\,$mm, a width of 3$\,$mm and a height of 0.5$\,$mm and was made of magnesium oxide doped PPLN. It was designed for SHG from 976$\,$nm to 488$\,$nm at a temperature around 100$\,^{\circ}$C and has 16 waveguide channels applied by reverse proton exchange \cite{Parameswaran2002}, each with a width of 5$\,\mu$m and a height of 3$\,\mu$m.

\begin{figure}[h]
\centerline{\includegraphics[width=9cm]{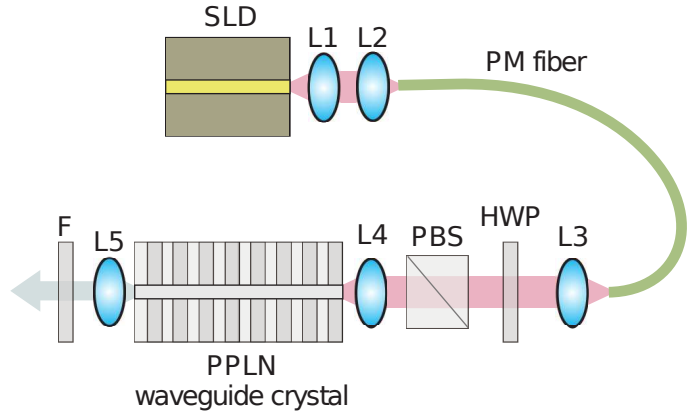}}
\caption{Sketch of the experimental setup. SLD - superluminescent diode, PM - single mode polarization maintaining fiber, HWP - half wave-plate, PBS - polarization beam splitter, L1-L5 aspherical lenses , F - spectral filters.}
\end{figure}

\section*{Experimental results}
The ASE of the BA SLD is characterized by a broadband multi-mode spectrum centered at 976\,nm and a spectral bandwidth of $\Delta \lambda \approx$10\,nm (FWHM) as shown in Fig. 2. Recently, it was shown that such ASE sources can possess chaotic or thermal-like photon statistics \cite{Boitier2009, Blazek2011}. Following Boitier et al. \cite{Boitier2009} we measured the DSOC of the emission from the SLD using a Michelson interferometer and a PMT as two-photon detector (\textsc{Becker $\&$ Hickl PMC100}). The resulting autocorrelation function was filtered and the DSOC was extracted as a function of the time delay. The extracted DSOC g$^{(2)}(\tau)$ is shown in Fig. 3. For zero time delay, a value of of g$^{(2)}$(0)\,=\,1.9$\pm$0.1 was determined, validating that the emission is incoherent. For details about the measurement technique and filtering procedure please see \cite{Boitier2009}. From the autocorrelation measurement, a coherence time of 95 fs (FWHM) was determined, which is in accordance with the measured spectral bandwidth of the ASE.

\begin{figure}[tp]
\centerline{\includegraphics[width=7cm]{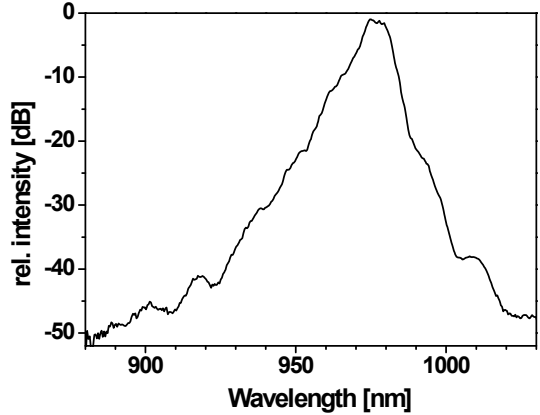}}
\caption{Spectrum of the ASE of the SLD at a maximum power coupled to the PM fiber of 800$\,\mu$W.}
\end{figure}
\begin{figure}[bp]
\centerline{\includegraphics[width=7cm]{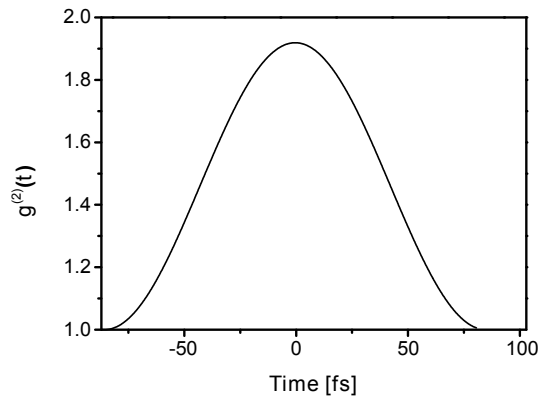}}
\caption{Measured DSOC of the ASE of the SLD at a maximum power coupled to the PM fiber of 800$\,\mu$W.}
\end{figure}

Figure 4 shows the SHG output power as a function of the coupled infrared pump power from the SLD emission. The injection current was held fixed and the light incident at the waveguide was attenuated using a half wave plate (HWP) in front of a polarizing beam splitter (PBS). The maximum power of the ASE with thermal-like photon statistics was 800$\,\mu$W coupled into the single mode fiber. As expected, at the low conversion limit, the SHG power follows the pump power quadratically: $P_{shg}=\eta_{shg} P_{fun}^2$ with the normalized conversion efficiency $\eta_{shg}$. A maximum SHG power of 23.5$\,$nW was obtained at 710$\,\mu$W of pump light coupled into the waveguide, yielding a conversion efficiency of about $\eta_{shg,ase}\approx 4.5\,\%/W$. This is in the range of normalized conversion efficiencies with PPLN bulk crystals \cite{JechowLPR2010, skoczowsky2010ol, Jensen2013} and laser radiation.

\begin{figure}[tp]
\centerline{\includegraphics[width=7cm]{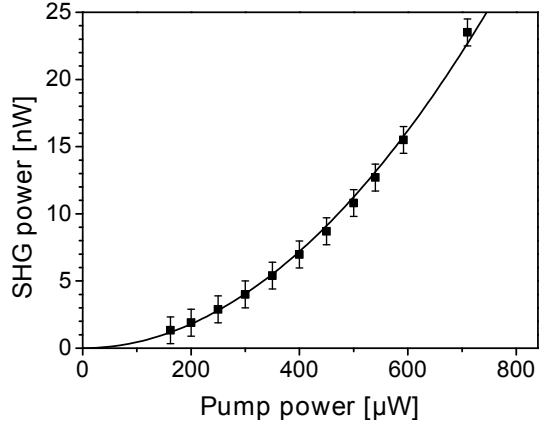}}
\caption{Power of the generated blue light as a function of the infrared pump power coupled into the PPLN waveguide crystal.}
\end{figure}
\begin{figure}[bp]
\centerline{\includegraphics[width=7cm]{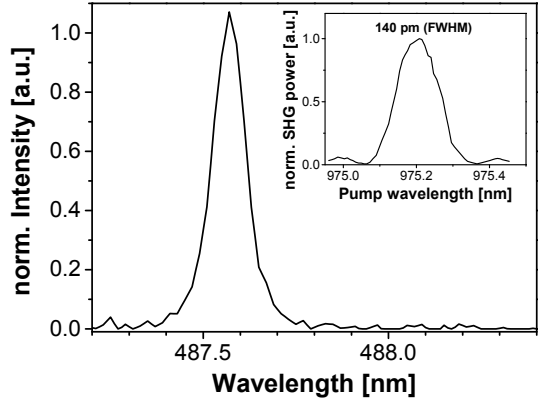}}
\caption{Spectrum of the frequency doubled ASE from the SLD at a waveguide temperature of 95$\,^{\circ}$C and maximum output power of the generated blue light. The inset shows the acceptance bandwidth measured with a tunable DFB diode laser.}
\end{figure}

The spectrum of the generated blue light measured with an optical spectrum analyzer (OSA) is shown in Fig. 5. The broad band ASE was converted to a narrow band spectrum with a width of 80$\,$pm (FWHM), dominated by the acceptance bandwidth of the PPLN crystal, which was measured to be $\approx$140$\,$pm in the infrared spectrum (see inset Fig. 5). For comparison we used a DFB diode laser under the same coupling conditions to pump the SHG inside the same waveguide channel. With 500$\,\mu$W coupled into the waveguide 750$\,$nW of blue light could be generated, giving a conversion efficiency of about $\eta_{shg,dfb}\approx 300\,\%/W$ in the same waveguide channel, which is in accordance to previous results at higher powers \cite{Jechow2007DFB, jechow07oe}.

We assume that the conversion efficiency is mainly limited by the acceptance band width of the crystal as investigated in \cite{Kontur2007}. In our case, the acceptance bandwidth of the crystal is 140$\,$pm and the bandwidth of the ASE source is about 10$\,$nm giving a ratio of $\Delta\lambda_{ASE} / \Delta\lambda_{PPLN} \approx 70$. The SHG conversion efficiency with the ASE source is reduced approximately by the same ratio compared to the DFB laser. By changing the temperature of the PPLN crystal, the QPM wavelength can be varied and the blue light can be tuned by 0.04$\,$nm/K. Figure 6 shows the SHG output power as a function of the crystal temperature for the ASE source (squares) and the DFB diode laser (dots), both operated at constant current. Due to the broad spectrum of the ASE it is possible to tune the blue light over a range of more than 4$\,$nm without simultaneously tuning the pump light source. In contrast to the stabilized, narrow bandwidth DFB laser the requirements of matching the pump laser wavelength to the QPM wavelength are drastically reduced. With the ASE source the SHG output power remains above 55$\%$ over a temperature range of 100$\,$K, while it drops below 10$\%$ for the  DFB diode laser at $\pm\,$1K.

\begin{figure}[h]
\centerline{\includegraphics[width=7cm]{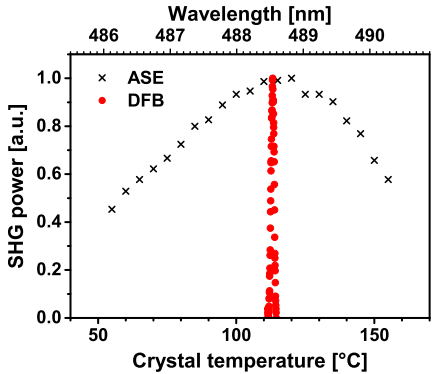}}
\caption{SHG output power as a function of the crystal temperature using thermal-like light (squares) and coherent light (dots).}
\end{figure}

\section*{Discussion and Conclusion}
We have used temporally incoherent emission from an SLD to perform SHG generating blue light easily visible by the naked eye. The ASE yielded a measured thermal-like photon statistics with a measured DSOC of g$^{(2)}$(0)\,=\,1.9$\pm$0.1. Although we were not able to measure the DSOC of the frequency doubled light, we want to point out that the bunching is possibly enhanced for the SHG light. According to Loudon (eq. 9.3.11) \cite{loudon}, the DSOC of the SHG light is given by:
\begin{equation}
g^{(2)}_{SHG}(0)= {g^{(4)}_{pump}(0) \over (g^{(2)}_{pump}(0))^2}.
\end{equation}
For coherent and thermal-like light the higher order DSOC are given by:
\begin{equation}
g^{(n)}_{coherent}(0) = 1 \hspace{5mm}and\hspace{5mm}  g^{(n)}_{thermal}(0) = n!
\end{equation}
which for SHG gives:
\begin{equation}
g^{(2)}_{SHG,coherent}(0) = 1 \hspace{5mm} and \hspace{5mm} g^{(2)}_{SHG,thermal}(0) = 6.
\end{equation}
This enhanced bunching would be of high interest for applications like microscopy \cite{jechow2013np} or ghost imaging \cite{Gatti2004} as well as fundamental quantum optical experiments.

Compared to early and recent results with incoherent light using birefringent phase matched material \cite{mcmahon1965, mcmahon1966, tamosauskas2011}, we achieve an SHG conversion efficiency that is six orders of magnitude higher, although it is difficult to compare the different light sources. This is achieved by using quasi-phasematched material and because of the single spatial mode properties of the ASE, which allowed coupling into a waveguide crystal. It has to be stated, that the PPLN waveguide crystal used here was explicitly designed for SHG with a narrow bandwidth pump source

For further optimization it should be possible to tailor the poling period in a way, that efficient SHG of broadband light can be realized \cite{Chou1999}. In principle, even a tailored SHG output spectrum of arbitrary shape can be generated by using such an engineered QPM material. Furthermore, ASE sources with higher output powers are available, which would increase the overall output power of the SHG light. ASE sources based on semiconductor diodes can be mass produced at low costs and because of the broadband emission the phase matching conditions are rather relaxed, making a precise temperature control of the crystal obsolete. This scheme might find application in e.g. upconversion imaging \cite{Dam2010} or photovoltaics \cite{zou2012}.

In conclusion, our results show that frequency doubling of light with thermal-like photon statistics is possible with reasonable efficiencies even at relatively low light levels and despite a non optimized scenario. The SHG light might find applications and is of interest for further studies in quantum optics. We outline optimization strategies that should result in higher efficiency and photon flux. To our knowledge, this is the first demonstration of SHG of ASE with known photon statistics.


\section*{Acknowledgments}
We thank Becker $\&$ Hickl GmbH for providing the PMT for the g$^{(2)}(\tau)$ measurement and Ralf Menzel for fruitful discussion. The authors declare no conflict of interest.

\section*{References}
\providecommand{\noopsort}[1]{}\providecommand{\singleletter}[1]{#1}%
\providecommand{\newblock}{}

\end{document}